\documentclass{emulateapj}

\usepackage{amssymb}
\usepackage{amsmath}
\usepackage{graphicx}
\usepackage{natbib}
\usepackage{txfonts}

\bibpunct{(}{)}{;}{a}{}{,} 

\def\roma{1}
\def\icra{2}

\shorttitle{Gravitational Waves versus X and Gamma Ray Emission in a Short Gamma-Ray Burst}
\shortauthors{Oliveira et al.}

\begin{document}
\title{Gravitational Waves versus X and Gamma Ray Emission in a Short Gamma-Ray Burst}

\author{F. G. Oliveira\altaffilmark{\roma,\icra},
				Jorge A. Rueda\altaffilmark{\roma,\icra},
				\&
        R. Ruffini\altaffilmark{\roma,\icra}
        }

\altaffiltext{\roma}{Dipartimento di Fisica and ICRA, 
                     Sapienza Universit\`a di Roma, 
                     P.le Aldo Moro 5, 
                     I--00185 Rome, 
                     Italy}

\altaffiltext{\icra}{ICRANet, 
                     P.zza della Repubblica 10, 
                     I--65122 Pescara, 
                     Italy}

\altaffiltext{}{fe.fisica@gmail.com,jorge.rueda@icra.it,ruffini@icra.it}

\begin{abstract}
The recent progress in the understanding the physical nature of neutron star equilibrium configurations and the first observational evidence of a genuinely short gamma-ray burst, GRB 090227B, allows to give an estimate of the gravitational waves versus the X and Gamma ray 
emission in a short gamma-ray burst.
\end{abstract}


\maketitle

\section{Introduction}

The first systematic analysis of the temporal distribution of the $T_{90}$, the observed duration of prompt emission of gamma-ray bursts (GRBs), made for the sample GRBs observed by the BATSE instrument on board the Compton Gamma-Ray Observer (CGRO) satellite \citep{1992Natur.355..143M}, evidenced a bimodal shape. This showed the existence of two different families of sources: \emph{long} and \emph{short} GRBs were defined as being longer or shorter than $T_{90} = 2$~s.

It has been recently that \cite{2013ApJ...763..125M} have probed the existence of the defined to as \emph{genuine} short GRB, theoretically predicted by the Fireshell model \citep{2001ApJ...555L.113R,2002ApJ...581L..19R} as bursts with the same inner engine as the long bursts but endowed with a severely low value of the baryon load, $B \equiv M_B c^2/E^{tot}_{e+e-}\lesssim 10^{-5}$. Here $M_B$ is the mass of the baryons engulfed by the expanding ultrarelativistic $e^+e^-$ plasma whose total energy is denoted by $E^{tot}_{e+e-}$; see section \ref{sec:3} for further details. The emission from these GRBs mainly consists in a first emission, the proper GRB (P-GRB), followed by a softer emission squeezed on the first one. The typical separation between the two components is expected to be shorter than 1--10 ms. No afterglow emission is expected from these sources.

Indeed, the time-resolved spectral analysis of the Fermi-GBM and Konus-Wind satellites data of GRB 090227B by \cite{2013ApJ...763..125M} has led to an estimate of the baryon load for this burst, $B=(4.13\pm 0.05)\times 10^{-5}$. The parameters inferred for GRB 090227B lead thus to the identification of the progenitor of the genuine short GRB in a neutron star binary \citep[see below in section~3 and][for details]{2013ApJ...763..125M}: (1) the natal kicks velocities imparted to a neutron star binary at birth can be even larger than 200~km~s$^{-1}$ and therefore a binary system can runaway to the halo of its host galaxy, clearly pointing to a very low average number density of the CBM; (2) the very large total energy, which we can indeed infer in view of the absence of beaming, and the very short time scale of emission point again to a neutron star binary; (3) as we shall show below in section~2 the very small value of the baryon load is strikingly consistent with two neutron stars having small crusts, in line with the recent neutron star theory \citep{2011PhLB..701..667R,2011NuPhA.872..286R,2012NuPhA.883....1B}.

The aim of this work is to make a detailed analysis of the neutron star binary progenitor of GRB 090227B. We compute the structure of the neutron star components following our recent model of neutron stars fulfilling global charge neutrality and including the strong, weak, electromagnetic, and gravitational interactions in the framework of general relativity and relativistic nuclear mean field theory. We simulate the evolution of the binary and compute the radiation emitted in form of gravitational waves that leads to the shrinking of the orbit and final merging. We compare and contrast the results of the dynamics as described by the classical test-mass limit approximation with the more accurate description based on the one-body formalism \citep{1999PhRvD..59h4006B,2000PhRvD..62f4015B,2000PhRvD..62h4011D,2001PhRvD..64l4013D,2010PhRvD..81h4016D}. We estimate the detectability of this kind of neutron star binaries by the Advanced LIGO interferometer. We compute also the total energy output in gravitational waves and compare it with the emission of the system in X and Gamma rays.

\section{Neutron star structure}\label{sec:2}

We have recently proved how the consistent treatment of neutron star equilibrium configurations, taking into account the strong, weak, electromagnetic, and gravitational interactions, implies the solution of the general relativistic Thomas-Fermi equations, coupled with the Einstein-Maxwell system of equations \citep{2011PhLB..701..667R,2011NuPhA.872..286R,2012NuPhA.883....1B}. This new Einstein-Maxwell-Thomas-Fermi (EMTF) equations supersede the traditional Tolman-Oppenheimer-Volkoff (TOV) equations, which impose the condition of local charge neutrality throughout the configuration \citep{tolman39,oppenheimer39}. We have shown that indeed this latter imposition of a TOV-like treatment explicitly violates the thermodynamic equilibrium of the star, which is ensured by the constancy of the generalized electro-chemical potentials (Klein potentials) of each system species along the whole configuration \citep{2011PhLB..701..667R,2011NuPhA.872..286R}.

The solution of the EMTF coupled differential equations introduces self-consistently the presence of the electromagnetic interactions in addition to the nuclear, weak, and gravitational interactions within the framework of general relativity. The weak interactions are accounted for by requesting the $\beta$-stability, and the strong interactions are modeled via the $\sigma$-$\omega$-$\rho$ nuclear model, where $\sigma$, $\omega$ and $\rho$ are the mediator massive vector mesons within relativistic mean field theory \'a la \cite{boguta77}. The nuclear model is fixed once the values of the coupling constants and the masses of the three mesons are fixed: in this work, as in the previous ones \citep{2012NuPhA.883....1B,belvedere14}, we adopt the NL3 parameter set \citep{lalazissis97}. The supranuclear core is composed by a degenerate gas of neutrons, protons, and electrons in $\beta$-equilibrium. The crust in its outer region $\rho \leq \rho_{\rm drip}\approx 4.3\times 10^{11}$ g~cm$^{-3}$ is composed ions and electrons and in its inner region, at $\rho_{\rm drip}<\rho<\rho_{\rm nuc}$, where $\rho_{\rm nuc}\approx 2.7\times 10^{14}$~g~cm$^{-3}$ is the nuclear saturation density, there is an additional component of free neutrons dripped out from nuclei.

\begin{figure}[!hbtp]
\includegraphics[width=\hsize,clip]{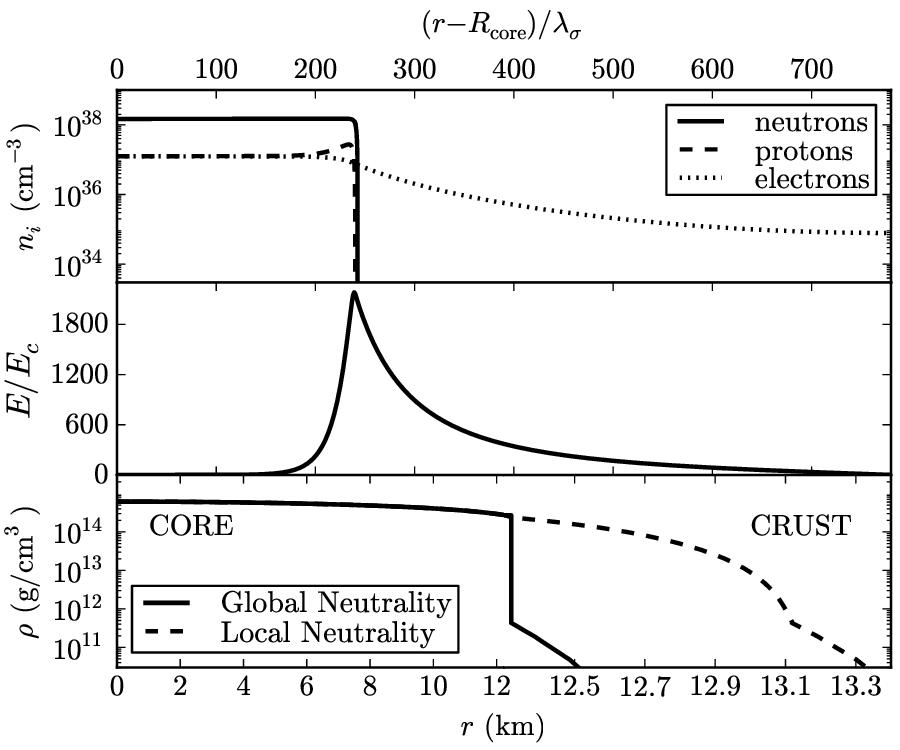}
\caption{Upper panel: particle density profiles in the core-crust boundary interface, in units of cm$^{-3}$. Middle panel: electric field in the core-crust transition layer, in units of the critical field $E_c$. Lower panel: density profile inside a neutron star with central density $\rho(0)\sim 5 \rho_{\rm nuc}$. We show here the differences between the solution obtained from the TOV equations (locally neutral case) and the globally neutral solution presented in \cite{2012NuPhA.883....1B}. In this example the density at the edge of the crust is $\rho_{\rm crust}=\rho_{\rm drip}=4.3\times 10^{11}$ g/cm$^3$ and $\lambda_\sigma = \hbar/(m_\sigma c) \sim 0.4$~fm denotes the $\sigma$-meson Compton wavelength.}\label{fig:profiles}
\end{figure}

The solution of the EMTF equations of equilibrium leads to a new structure of the neutron stars very different from the traditional configurations obtained through the TOV equations (see Fig.~\ref{fig:profiles}): the core is positively charged as a consequence of the balance between gravitational and Coulomb forces that results in the appearance of a Coulomb potential energy $e V\sim m_\pi c^2$ deep. The core-crust transition starts at $\rho=\rho_{\rm nuc}$. The transition is marked by the existence of a a thin, $\Delta r\sim$few hundreds fm, electron layer fully screening the core charge. In this transition layer the electric field becomes overcritical, $E\sim m^2_\pi c^3/(e \hbar)$, and the particle densities decrease until the base of the crust, which is reached when global charge neutrality is achieved. Consequently, the core is matched to the crust at a density $\rho_{\rm crust}\leq \rho_{\rm nuc}$. 

\begin{figure}[!hbtp]
\centering\includegraphics[width=\hsize,clip]{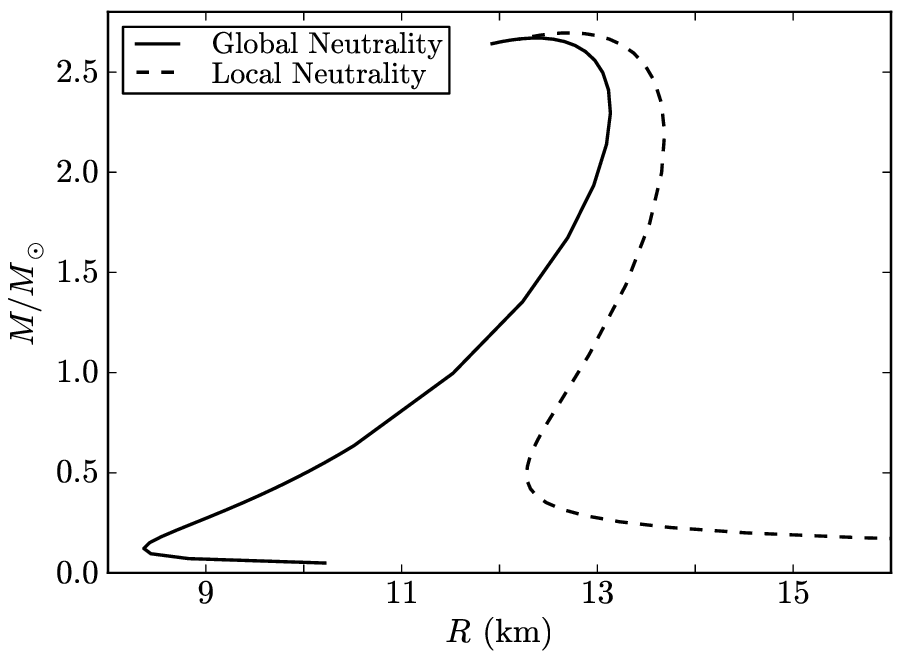}
\caption{Mass-radius relation obtained with the traditional locally neutral TOV treatment case and the global charge neutrality configurations, with $\rho_{\rm crust}=\rho_{\rm drip}$ \citep{2012NuPhA.883....1B}. Configurations lying between the solid and dashed curves have $\rho_{\rm crust} >\rho_{\rm drip}$ and so they possess inner crust.}
\label{fig:MR}
\end{figure}

For each central density there exists an entire family of core-crust interface boundaries and, correspondingly, a family of crusts with different mass $M_{\rm crust}$ and thickness $\Delta R_{\rm crust}$. The larger $\rho_{\rm crust}$, the smaller the thickness of the core-crust interface, the peak of the electric field, and the larger the $M_{\rm crust}$ and $\Delta R_{\rm crust}$. Configurations with $\rho_{\rm crust}>\rho_{\rm drip}$ possess both inner and outer crust while in the cases with $\rho_{\rm crust}\leq \rho_{\rm drip}$ the neutron star have only outer crust. In the limit $\rho_{\rm crust}\to\rho_{\rm nuc}$, both $\Delta r$ and $E$ of the transition layer vanish, and the solution approaches the one given by local charge neutrality \citep[see Figs.~3 and 5 in][]{2012NuPhA.883....1B}). All the above features lead to a new mass-radius relation of neutron stars; see \cite{2012NuPhA.883....1B} and Fig.~\ref{fig:MR}. The extension to the case of uniformly rotating neutron stars has been recently achieved in \citep{belvedere14}.

It is also worth to mention that in \citep{2012NuPhA.883....1B} we showed the agreement of this new mass-radius relation with the most stringent observational constraints to the mass-radius relation of neutron stars, that are provided by the largest mass, the largest radius, the highest rotational frequency, and the maximum surface gravity, observed from pulsars \citep{trumper2011}. They are imposed by the mass of PSR J1614-2230 $M=1.97 \pm 0.04 M_\odot$ \citep{demorest2010}, the lower limit to the radius of RX J1856-3754 \citep{trumper04}, the 716 Hz PSR J1748-2246ad \citep{hessels06}, and the surface gravity of the neutron star in the low-mass X-ray binary X7 from which 90\% confidence level contours of constant $R_\infty$ can be extracted \citep{heinke06}. 

It is known that uncertainties in the behavior of the nuclear matter at densities larger than the nuclear saturation density $n_{\rm nuc}\approx 0.16$ fm$^{-3}$, reached in the core of a neutron star, lead to a variety of nuclear equations of state which lead to different numerical estimates of the neutron star parameters, in particular the mass-radius relation. However, as we have mentioned above the current observational constraints strongly favor stiff nuclear equations of state such as the ones obtained from relativistic mean field models \'a la \cite{boguta77} used here, which provide high values for the critical mass of the neutron star, larger than the mass of PSR J1614-2230. This has been recently reconfirmed by the measurements of the high mass of PSR J0348+0432, $M=2.01 \pm 0.04~M_\odot$ \citep{antoniadis13}.

In Table \ref{tab:MRcrit} we show the critical mass and corresponding radius of globally neutral neutron stars obtained for selected parameterizations of the nuclear model used in \citep{2012NuPhA.883....1B}: NL3 \citep{lalazissis97} NL-SH \citep{sharma93}, TM1 \citep{sugahara94}, and TM2 \citep{hirata95}.

\begin{deluxetable}{lcccc}[!hbtp]
\tabletypesize{\scriptsize} 
\tablecaption{Critical mass and corresponding radius of globally neutral neutron stars for selected nuclear equations of state.
\label{tab:MRcrit}}
\tablewidth{0pt}
\tablehead{
\colhead{} &
\colhead{NL3} &
\colhead{NL-SH} &
\colhead{TM1} &
\colhead{TM2} 
}
\startdata
$M_{\rm crit}$ ($M_\odot$)&  2.67  & 2.68  & 2.58 & 2.82 \\
$R$ (km) & 12.33  & 12.54 & 12.31 & 13.28
\enddata
\end{deluxetable}
%

\section{Parameters of GRB 090227B}\label{sec:3}

In parallel, the theoretical progress in the Fireshell model of GRBs \citep[see]{2001ApJ...555L.107R,2001ApJ...555L.113R,2001ApJ...555L.117R} led to an alternative explanation of the Norris-Bonnell sources as disguised short burst \citep{2007A&A...474L..13B,2008AIPC..966....7B,2009A&A...498..501C,2010A&A...521A..80C,2011A&A...529A.130D}: canonical long bursts exploding in halos
of their host galaxies, with an average value of the circumburst (CBM) medium density $\langle n_{CBM} \rangle \approx 10^{-3}$ particles/cm$^3$. 

We now turn to the analysis of GRB 090227B. We first recall that the canonical GRB within the Fireshell model has two components: an emission occurring at the transparency of the optically thick expanding $e^+e^-$ baryon plasma \citep{2000A&A...359..855R}, the Proper-GRB (P-GRB), followed by the extended afterglow, due to the interactions between the accelerated baryons and the circumburst medium (CBM) of average density $\langle n_{CBM} \rangle$. Such an extended afterglow comprises the prompt emission as well as the late phase of the afterglow \citep{2005ApJ...620L..23B,2005ApJ...633L..13B}. The relative energy of these two components, for a given total energy of the plasma $E^{tot}_{e+e-}=E^{GRB}_{tot}$, where $E^{GRB}_{tot}$ is the observed GRB energy, is uniquely a function of the baryon load $B = M_B c^2/E^{GRB}_{tot}=M_B c^2/E^{tot}_{e+e-}$, with $M_B$ the mass of the baryons engulfed by the expanding ultrarelativistic $e^+e^-$ plasma. 

As we mentioned, an extremely low value of the baryon load, $B\lesssim 10^{-5}$ together with a low density of the CBM lead to a genuinely short GRB emission, in which no afterglow emission is observed. This is indeed the case of GRB 090227B.

From the 16 ms time-binned light curves a significant thermal emission in the first 96 ms, which has been identified with the P-GRB, has been found \cite{2013ApJ...763..125M}. The subsequent emission is identified with the extended afterglow. The P-GRB of 090227B has the highest temperature ever observed, $k_B T = 517$ keV, where $k_B$ is the Boltzmann constant. The results of the fit of the light curve and spectrum of GRB 090227B are summarized in Table \ref{tab:GRB090227B}. In particular we show the total energy emitted $E^{GRB}_{tot}$, Baryon load $B$, Lorentz factor at transparency $\Gamma_{tr}$, cosmological redshift $z$, intrinsic duration of the GRB emission $\Delta t$, and average density of the CBM $\langle n_{CBM}\rangle$; we refer to \cite{2013ApJ...763..125M} for further details.

\begin{deluxetable}{lccccc}[!hbtp]
\tabletypesize{\scriptsize} 
\tablecaption{Properties of GRB 090227B. $E^{GRB}_{tot}$ is the total energy emitted in the GRB, $B$ is the baryon load, $\Gamma_{tr}$ is the Lorentz factor at transparency, the cosmological redshift is denoted by $z$, the intrinsic duration of the GRB is $Delta t$, and the average density of the CBM is $\langle n_{CBM}\rangle$. We refer to \cite{2013ApJ...763..125M} for additional details.
\label{tab:GRB090227B}}
\tablewidth{0pt}
\tablehead{
\colhead{$E^{\rm GRB}_{\rm tot}$ (erg)} &
\colhead{$B$ } &
\colhead{$\Gamma_{tr}$} &
\colhead{$z$ } &
\colhead{$\Delta t$ (s)} &
\colhead{$\langle n_{\rm CBM}\rangle$ (cm$^{-3}$)}
 }
\startdata
 $2.83\times 10^{53}$   &   $4.13\times 10^{-5}$   & $1.44\times 10^4$  & $1.61$  & $0.35$ & $1.9\times 10^{-5}$
\enddata
\end{deluxetable}

\section{Inference of neutron star binary parameters}\label{sec:4}

We now infer the binary component parameters. It is clear that the merging of two neutron stars will lead to a GRB if the total mass of the binary satisfies
\begin{equation}\label{eq:Mcrit}
M_1+M_2 \gtrsim M_{\rm crit} = 2.67~M_\odot\, ,
\end{equation}
where $M_{\rm crit}$ is the critical mass over which a neutron star undergoes gravitational collapse to a black hole. For the numerical estimates we adopt the neutron star configurations obtained with the NL3 parameterization of the nuclear model (see Table \ref{tab:MRcrit}).

Assuming for simplicity a binary with twin components $M_1=M_2=M$, we obtain masses $M = 1.335~M_\odot$ and correspondingly radii $R_1=R_2=12.24$~km (see Fig.~\ref{fig:MR} and \cite{2012NuPhA.883....1B}). The mass of the corresponding crust of each component is $M_{\rm crust}\approx 3.6\times 10^{-5}~M_\odot$ and the thickness of the crust is $\Delta R_{\rm crust} \sim 0.47$ km. For the other nuclear parameterizations of the nuclear model the maximum stable mass and corresponding radius are given by Table \ref{tab:MRcrit}, and consequently the parameters of the single neutron star components change accordingly if described by such models, including the estimate of the baryon load, as shown below.

The location of the binary in the very low interstellar density medium of galactic halos makes possible to probe the neutron star theory and equation of state through the knowledge of the baryon load $B$ inferred from the fitting of the GRB light curve and spectrum. The baryonic matter which the GRB interact with is in these systems provided by the material of the neutron star crusts ejected during the binary coalescence. Thus, a theoretical expectation of the baryon load $B$ left in a binary neutron star merger is
\begin{equation}\label{eq:Bth}
B_{\rm th}=\eta \frac{M^{tot}_{\rm crust}c^2}{E^{GRB}_{tot}}\, ,
\end{equation}
where $\eta$ is the fraction of the total crustal mass, $M^{tot}_{\rm crust}=M_{1,\rm crust}+M_{2,\rm crust}=2 M_{\rm crust}=7.2\times 10^{-5}~M_\odot$, which is ejected. We assume that the mass ejected during the merger comes from the the crust of the two neutron star components of the system, as should be expected from a symmetric binary merger.

\begin{figure}[!hbtp]
\includegraphics[width=\hsize,clip]{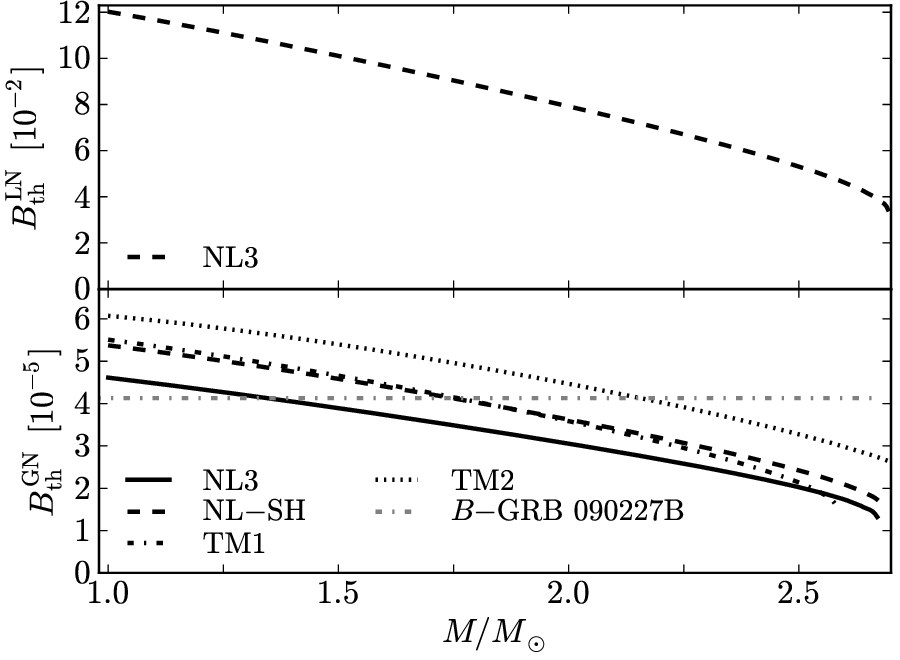}
\caption{Theoretical baryon load expected to be left by a binary neutron star merger as given by Eq.~(\ref{eq:Bth}), for $\eta=0.1$, as a function of the total mass $M$. Upper panel: locally neutral neutron stars, $B^{\rm LN}_{\rm th}$, for the NL3 parameterization of the nuclear model, dashed black curve, units $10^{-2}$. Lower panel: globally neutral neutron stars, $B^{\rm GN}_{\rm th}$, for the NL3, NL-SH, TM1, and TM2 parameterizations of the nuclear model. We indicate the observed baryon load of GRB 090227B, $B=4.13\times 10^{-5}$, with the dashed-dotted gray horizontal line; see Table \ref{tab:GRB090227B} and \cite{2013ApJ...763..125M}.}\label{fig:crust}
\end{figure}

In Fig.~\ref{fig:crust} we have plotted the theoretical baryon load given by Eq.~(\ref{eq:Bth}) for GRB 090227B, namely using $E^{GRB}_{tot}=2.83\times 10^{53}$ erg, as a function of the mass $M$ of the neutron star for the nuclear equations of state of Table \ref{tab:MRcrit}. For the locally neutral case we use for the sake of comparison only the result for the NL3 parameterization.

The agreement of the observed baryon load of GRB 090227B (see Table \ref{tab:GRB090227B} and \cite{2013ApJ...763..125M}) with the low mass of the crust obtained from the globally neutral neutron stars of \cite{2012NuPhA.883....1B} is evident (see Fig.~\ref{fig:crust}). It can be compared and contrasted with the ones obtained enforcing the local charge neutrality condition. For the specific binary neutron star system adopted here, we obtain a theoretical prediction of the baryon load from Eq.~(\ref{eq:Bth}) with $\eta=1$, $B_{\rm th}\approx 4.5\times 10^{-4}$, or a mass of the baryons $M_B = M^{tot}_{\rm crust}\approx 7.2\times 10^{-5}~M_\odot$, to be confronted with the one obtained from the fitting procedure of GRB 090227B, $B \sim 4.13\times 10^{-5}$, corresponding to $M_B = B\times E^{GRB}_{tot}/c^2 \sim 0.7\times 10^{-5}~M_\odot$. A perfect agreement requires $\eta\approx 0.1$ for the NL3 nuclear model, while the other nuclear parameterizations require a slightly lower value of $\eta$; see Fig.~\ref{fig:crust}. The above theoretical predictions of the neutron star crust mass $M_{\rm crust}$ and consequently the value of $E^B_{\rm crust}$ and $B$ have been inferred for a crust with a density at its edge equal to the neutron drip density $\rho_{\rm drip}\sim 4.3\times 10^{11}$~g~cm$^{-3}$. Neutron star crusts with densities $\rho<\rho_{\rm drip}$ are predicted by the new neutron star theory \citep{2012NuPhA.883....1B}, therefore there is still room for smaller values of the baryonic matter ejected in a binary process, and consequently to still shorter genuinely short GRBs.

The mass-energy of the baryon ejecta obtained from the estimate (\ref{eq:Bth}) gives for locally neutral neutron stars values $10^2$--$10^3$ bigger than the ones analyzed before (see Fig.~\ref{fig:crust}), due to the more massive crusts obtained from the TOV-like treatment \citep[see][for details]{2012NuPhA.883....1B}. It implies that Eq.~(\ref{eq:Bth}) gives in such a case $M_B\sim 10^{-3}$--$10^{-2}~M_\odot$, in line with previous results obtained from the numerical simulation of the dynamical evolution of neutron star binaries \citep[see e.g.][]{2001A&A...380..544R,2011A&A...531A..78G}, where locally neutral neutron stars are employed.

Turning to a possible alternative scenario, the crust supported by strange quark stars has densities strictly lower than $\rho_{\rm drip}$ and therefore they have crust masses $\sim 10^{-5}~M_\odot$ \citep[see e.g.][for details]{alcock86,1992ApJ...400..647G}, similar to the crust of the globally neutron neutron stars \citep{2012NuPhA.883....1B}. This leads to the natural question whether strange stars could be also a viable explanation to the low value of the baryon load of short GRBs. In addition, the quark core-crust transition is also characterized by overcritical electric fields. However, the softness of the equation of state of strange quark matter leads to a mass-radius relation for these stars characterized by a low maximum stable mass and small radii, ruled out by the current observational constraints of pulsars which put a lower limit to the radius of a compact star with $M = 1.4~M_\odot$, $R\gtrsim 12$~km \citep{trumper2011}, and the most massive compact star observed, PSR J0348+0432 with $M=2.01 \pm 0.04~M_\odot$ \citep{antoniadis13}.

\section{Gravitational wave emission}\label{sec:5}

The emission of gravitational waves signals from binaries system are the most expected signals to be detect by the interferometers called Advanced LIGO\footnote{http://www.advancedligo.mit.edu}-VIRGO\footnote{http://www.cascina.virgo.inft.it} and they have been planned for to be operational in a few years with a improved sensitivity approximately a factor of 10 better than the first generation of detectors. The connection between short gamma-ray signals and gravitational waves signals as a coincidence of the same event would allow in principle to understand more about the origin of short GRBs \citep[see][and references therein]{2003ApJ...589..861K}. 

We use here the adiabatic approximation to estimate the gravitational wave emission from the binary neutrons star. We used the above values of the neutron star binary progenitor estimated for the short GRB 090227B at a cosmological redshift $z=1.61$ \citep{2013ApJ...763..125M}. We adopt for simplicity circular orbits. First we compute the dynamics following the classic non-relativistic test-mass limit approximation and compare it with the more accurate description given by the one-body formalism, which accounts for the effects of general relativity.

\subsection{Classical Dynamics}\label{sec:5.1}

The orbital angular velocity of the binary with components $(M_1,R_1)$ and $(M_2,R_2)$ orbiting each other in a circular orbit of radius $r$, is given by 
\begin{equation}
\Omega = \sqrt{\frac{G (M_1 + M_2)}{r^3}}\, ,
\end{equation}
and its total binding energy is
\begin{equation}\label{eq:Etot}
E_b=-\frac{1}{2} \frac{G M_1 M_2}{r}\, .
\end{equation}
The leading term driving the loss of binding energy via gravitational wave emission is given by
\begin{equation}\label{eq:Ebdot}
-\frac{d E_b}{dt} = \frac{32}{5} \frac{G^4}{c^5} \frac{(M_1+M_2) (M_1 M_2)^2}{r^5}\, ,
\end{equation}
which leads to a decreasing of the separation $r$ with time and consequently a shortening of the orbital period $P=2\pi/\Omega$ dictated by \citep{landaubook}
\begin{equation}
\frac{1}{P} \frac{dP}{dt} = \frac{3}{2}\frac{1}{r} \frac{dr}{dt} = -\frac{3}{2} \frac{1}{E_b} \frac{dE_b}{dt}.
\end{equation}

The loss of orbital binding energy by emission of gravitational waves from the neutron star system in spiral phase for non-relativistic and point-like particles can be written as a function of the gravitational waves frequency $f$ as
\begin{equation}
 \frac{dE_b}{df} = -\frac{1}{3} (\pi G)^{2/3} {\cal M}^{5/3} f^{-1/3}, \label{classical}
\end{equation}
where ${\cal M} = (M_1M_2)^{3/5}/(M_1+M_2)^{1/5}$ is the called chirp mass.

\subsection{Effective one-body dynamics}\label{sec:5.2}

\begin{figure}[!hbtp]
\includegraphics[width=\hsize,clip]{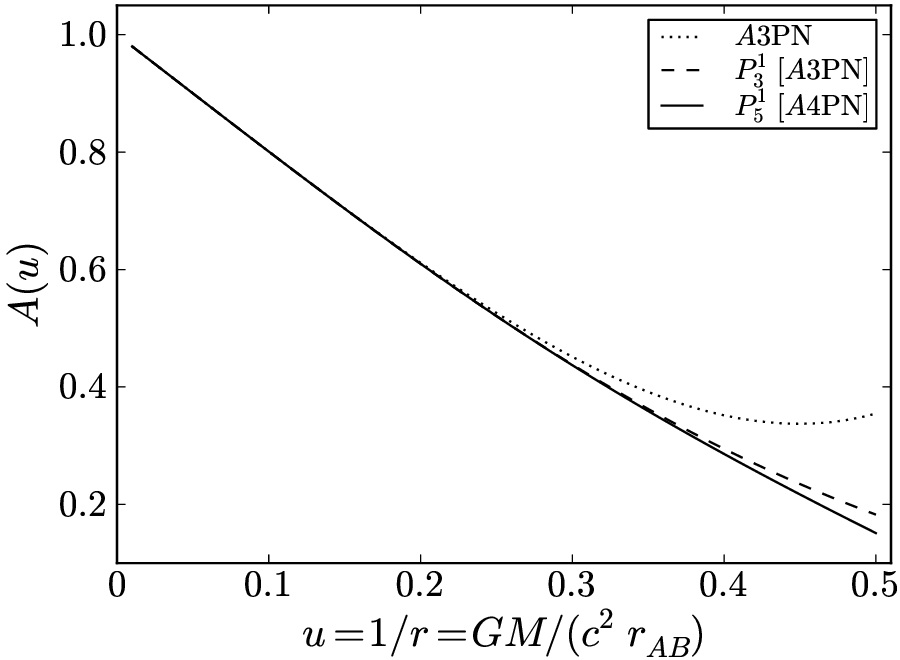}
\caption{Comparison between the EOB radial potential as a function of the u-parameter for the cases: $A(u;\nu)=3{\rm PN}$ (dotted), $P^1_3[A(u;\nu)= 3{\rm PN}]$ (dashed) and $P^1_5[A(u;\nu)=4{\rm PN}]$ (solid), where the $P^m_n [.]$ is the Pad\`e approximant.}
\label{Axu}
\end{figure}

The effective one-body (EOB) formalism \citep{1999PhRvD..59h4006B,2000PhRvD..62f4015B,2000PhRvD..62h4011D,2001PhRvD..64l4013D,2010PhRvD..81h4016D} maps the conservative dynamics of a binary system of non spinning objects onto the geodesic dynamics of one body of reduced mass $\mu=M_1 M_2/M$, with $M=M_1+M_2$ the total binary mass. The effective metric is a modified Schwarzschild metric given by
\begin{equation}
 ds^2 = -A(r)dt^2 + B(r)dr^2+ r^2(d\theta^2+sin^2 \theta d\phi^2)
\end{equation}
where the rescaled radial coordinate $r = c^2 r_{12}/(G M)$ has been introduced, with $r_{12}$ the distance between the two stars. The radial potential is given by
\begin{equation}
 A(u;\nu) = 1 - 2u + 2 \nu u^3 + a_4 \nu u^4 +a_5 \nu u^5,
\end{equation}
where $u = 1/r=G M/(c^2 r_{12})$, $\nu={M_1M_2}/{(M_1+M_2)^2}$ is the symmetric mass ratio (see Fig.~\ref{Axu}), with the values of the 3 and 4 post-Newtonian (PN)-level coefficients given by $a_4 = 94/3 - (41/32)\pi^2$ and $a_5(\nu) = a5^{c0} + \nu a5^{cl}$ \citep[see][for details]{2013PhRvD..87l1501B}. We will denote to as $P^m_n$ the Pad\`e approximant of order $(n,m)$, which when applied to $A(u;\nu)$ ensures the convergence of the solution near the merger point \citep[see][and references therein]{2009PhRvD..79h1503D}.

The EOB Hamiltonian is
\begin{equation}
 H = M c^2\sqrt{1 + 2 \nu (\hat{H}_{\rm eff}-1)},
\end{equation}
and effective Hamiltonian is described by
 \begin{equation}
  \hat{H}_{\rm eff}^2  = A(u) + p_{\phi}^2 B(u),
 \end{equation}
where $B(u) = u^2 A(u)$ and the angular momentum for the circular orbit is given by
\begin{equation}
 p_{\phi}^2 = - \frac{A'(u)}{[u^2 A(u)]'},
\end{equation}
where prime stands for derivative with respect to $u$.

We need to write $\hat{H}_{\rm eff}$ as a function of the orbital angular velocity $\Omega$, or orbital frequency $f$. For this, we need to write the $u$-parameter as a function of $\Omega$, or $f$, which is obtained from the angular Hamilton equation of motion in the circular case
\begin{equation}
 G M\Omega(u) = \frac{1}{u} \frac{\partial H}{\partial p_{\phi}} = \frac{M A(u) p_{\phi}(u) u^2}{H \hat{H}_{\rm eff}}. \label{EBO}
\end{equation}

\begin{figure}[!hbtp]
\includegraphics[width=\hsize,clip]{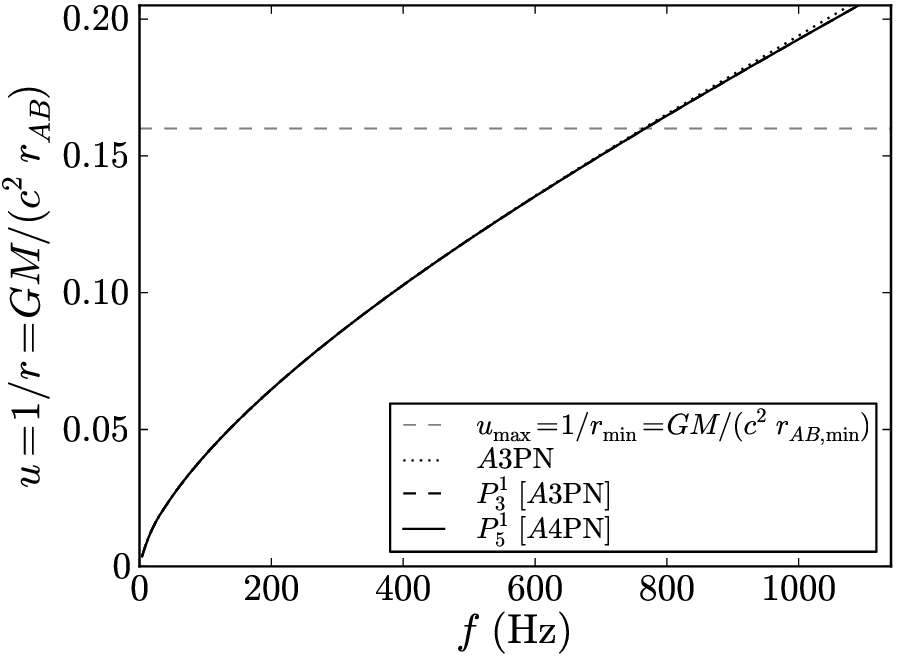}
\caption{The $u$ parameter as a function of the source frequency obtained from Eq.~\ref{EBO} in the case of a symmetric binary, $M_1=M_2$, so $\nu=1/4$. Here $r_{AB,{\rm min}}=R_1+R_2$ is the end point of the adiabatic approximation of the EOB formalism, which in this case is the point where the two neutron stars touch each other.}
\label{uvsom}
\end{figure}

We follow the binary evolution up to the \emph{contact} orbital frequency, $\Omega_c$, which we compute at the location of the innermost stable circular orbit, $R_{\rm ISCO}$, or when the two stars effectively \emph{touch} each other, namely at $r_{AB,{\rm min}}=R_1+R_2$, if $R_{\rm ISCO}<R$. Within the EOB formalism, the value of $R_{\rm ISCO}$ is given by the solution of the equation
\begin{equation}
A'(u_{\rm ISCO}) B''(u_{\rm ISCO})-A''(u_{\rm ISCO}) B'(u_{\rm ISCO})=0,
\end{equation}
which for our binary neutron star gives $u_{\rm ISCO}=[0.25,0.24,0.2,0.2]$ for $A(u)=1-2 u$ (test-mass limit dynamics, $R_{\rm ISCO}=6 G M/c^2$), $A(u;\nu)=3{\rm PN}$, $P^1_3[A(u)= 3{\rm PN}]$, and $P^1_5[A(u;\nu)=4{\rm PN}]$, respectively. Since $u_{\rm max}=G M/(c^2 r_{AB,{\rm min}})=0.16<u_{\rm ISCO}$, our contact point is given by $u_{\rm max}$ and not by $u_{\rm ISCO}$. Therefore, in the present case the contact orbital frequency is given by $\Omega_c=\Omega(u_{\rm max})$. In Fig.~\ref{uvsom} we show the result of the numerical integration of the above equation for the present binary system.

\begin{figure}[!hbtp]
\includegraphics[width=\hsize,clip]{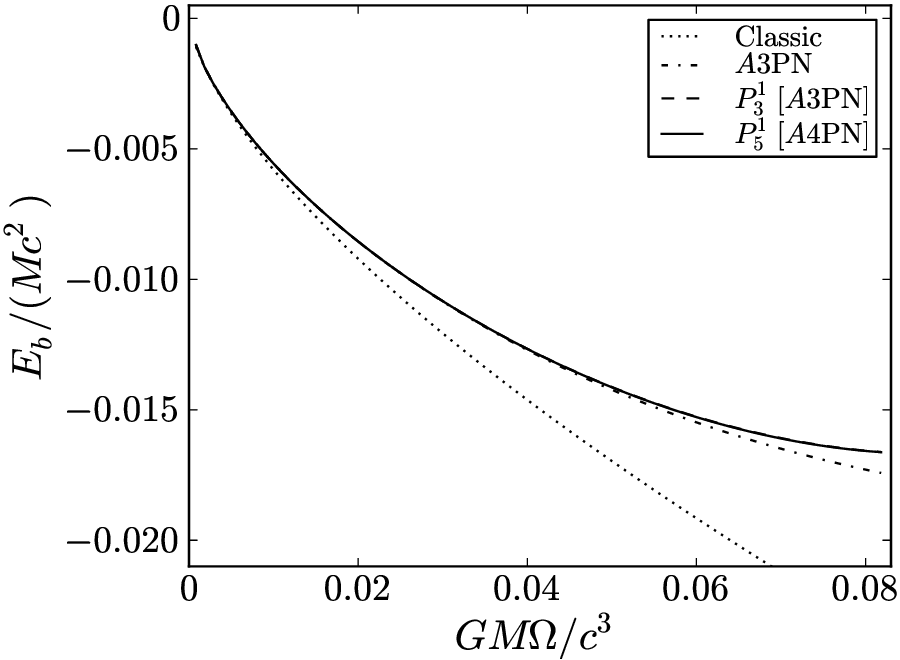}
\caption{Comparison of the EOB binding energies using the radial potential $\rm A(u;\nu)=3PN$, $P^1_3[\rm A(u;\nu)= 3PN]$ and $P^1_5[\rm A(u;\nu)=4PN]$ and the classic dynamics.}
\label{Eb}
\end{figure}

The binding energy as a function of the orbital frequency (see Fig.~\ref{Eb}) is,
\begin{equation}
 E_b(\Omega) = H - M c^2 = M c^2 [\sqrt{1 + 2 \nu (\hat{H}_{\rm eff} - 1})-1 ],
\end{equation}
and the gravitational energy spectrum is obtained through the derivative $dE_b/d\Omega$.

\section{Detectability: signal-to-noise ratio}\label{sec:6}

A positive detection of the gravitational waves emission implies that the signal overcomes some minimum threshold value of the signal-to-noise ratio (SNR). The SNR appropriate for matched-filtering search is given by \citep[see][for details]{1998PhRvD..57.4535F}
\begin{equation}\label{eq:rho1}
 {\rm SNR}^2 = 4 \int_0^{\infty}\frac{|\tilde{h}(f)|^2}{S^2_h(f)} df,
\end{equation}
where $\tilde{h}(f)$ is the Fourier transform of the gravitational waveform $h(t)$ and $S_h(f)$ is the strain noise spectral density (units 1/$\sqrt{{\rm Hz}}$) in the interferometer. Besides its dependence on the waveform, the above SNR depends in general on the orientation and position of the source with respect to the interferometer.

After making an rms average over all the possible source orientations, positions, and wave polarizations, the SNR given by Eq.~(\ref{eq:rho1}) becomes
\begin{equation}\label{eq:SNR}
\langle{\rm SNR}^2\rangle=\int_{f_{\rm min}}^{f_{\rm max}} d f_d \frac{h^2_c(f_d)}{5 f^2_d S^2_h(f_d)},
\end{equation}
where we have introduced the characteristic gravitational waves amplitude, $h_c$, defined using the Fourier transform of the gravitational waveform $h(t)$, $h_c(f)=f|\tilde{h}(f)|$, and it is given by
\begin{equation}\label{hc}
 h_c^2(f)= \frac{2(1+z)^2}{\pi^2 d_L^2}\frac{G}{c^3} \frac{d E_b}{d f} [(1+z) f_d].
\end{equation}
with $z$ the cosmological redshift, $f_d= f/(1+z)$ the gravitational wave frequency at the detector, $f=\Omega/\pi$ the frequency in the source frame, $\Omega$ is the orbital frequency, the minimal bandwidth frequency of the detector is $f_{\rm min}$, and $f_{\rm max}=f_c/(1+z)$ is the maximal bandwidth frequency, where $f_c=\Omega_c/\pi$ is the binary contact frequency. We use in this work a standard cosmological model with $H_0=75$~km/s/Mpc, $\Omega_M=0.27$, and $\Omega_\Lambda=0.73$, and a luminosity distance $d_L(z)=(c/H_0)(1+z)\int_0^z [\Omega_M (1+x)^3+\Omega_\Lambda]^{-1/2} dx$. We adopt int his work as threshold value for a positive detection, $\langle {\rm SNR}\rangle=5$, following previous works \citep[see e.g.][]{2003ApJ...589..861K}. 

The rms averaged SNR given by Eq.~(\ref{eq:SNR}) depends only on the distance to the source, i.e. the cosmological redshift, and the energy-spectrum, $dE_b/df$, of the gravitational waves; we refer the reader to \citep{1998PhRvD..57.4535F} for further details on the rms averaging. 

In order to assess the detectability of the source in the spiraling-in phase, the integration in Eq.~(\ref{eq:SNR}) is carried out from the minimum bandwidth frequency of the interferometer all the way up to the contact frequency, namely the maximum frequency which is given by the merger point, namely when the two stars touch each other.

We now compare and contrast the characteristic amplitude per square root frequency, $h_c(f_d)/\sqrt{f_d}$, with the the strain noise spectral density $S_h(f)$ of the Advanced LIGO interferometer, as a function of the frequency at the detector, $f_d$. In this work we use the \emph{Optimal NSNS} noise curve of the LIGO Document T0900288-v3, which is optimized for a $1.4~M_\odot$ neutron star, and gives a $\langle {\rm SNR}\rangle=8$ at a distance of $200$~Mpc for a single interferometer\footnote{https://dcc.ligo.org/LIGO-T0900288/public - The curve represent the incoherent sum of the principal noise sources best understood at this time, namely the quantum, seismic, and thermal noises. There will be, in addition, technical noise sources. This is not a guaranteed performance, but a good guide to the overall curve and an early approximation to the anticipated sensitivity reachable by Advanced LIGO.}. The comparison is made for both the dynamics given by the non-relativistic point-like particles approximation (see section \ref{sec:5.1}) and the dynamics obtained from the EOB formalism (see section \ref{sec:5.2}).

\begin{figure*}[!hbtp]
\includegraphics[width=0.49\hsize,clip]{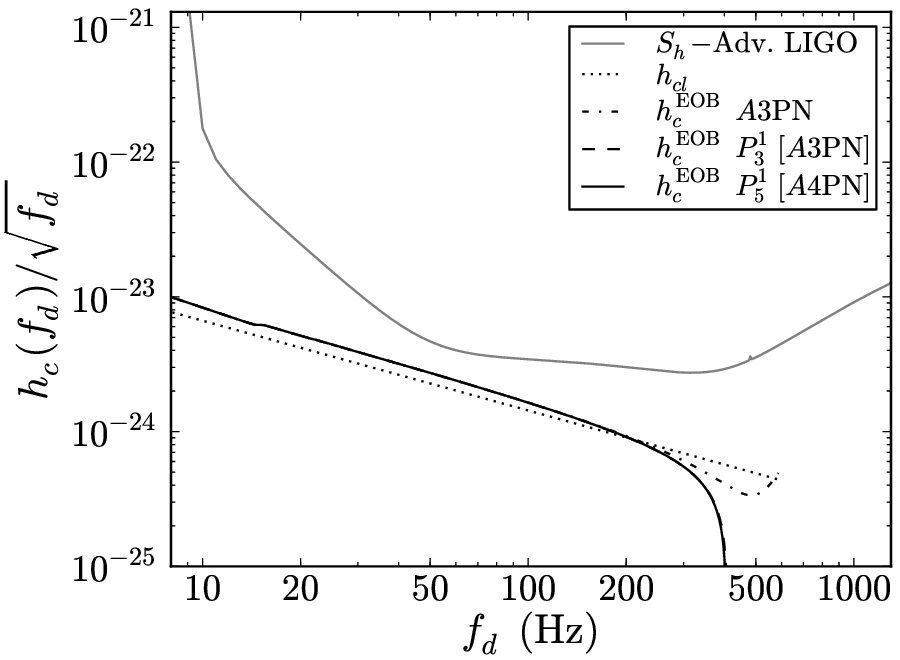}\includegraphics[width=0.49\hsize,clip]{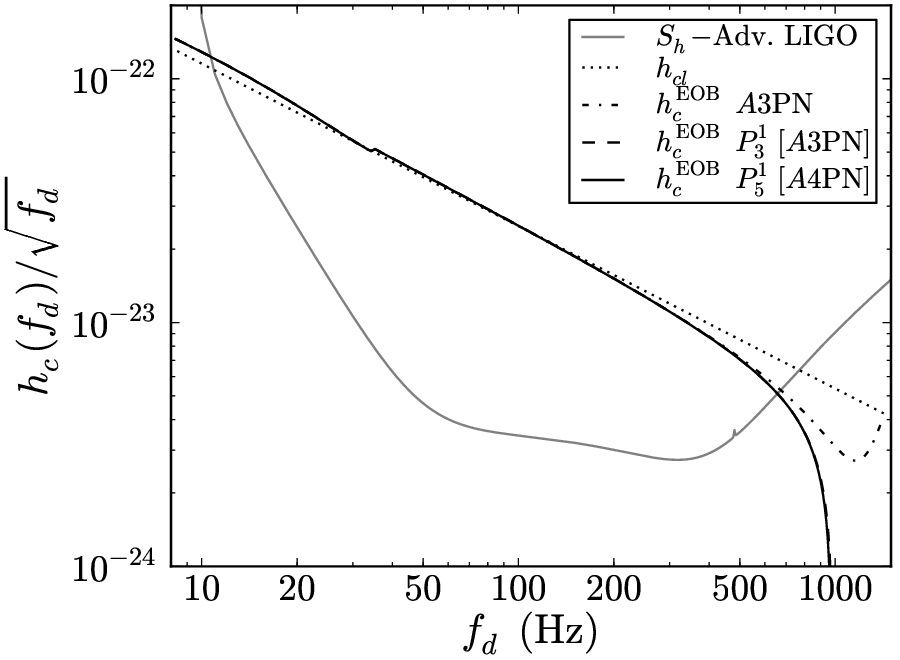}
\caption{Left panel: Comparison of the characteristic gravitational waves amplitude per unit square root frequency, $h_c(f_d)/\sqrt{f_d}$, see Eq.~(\ref{hc}), with the noise density spectrum $S_h(f)$ of the Advanced LIGO interferometer. We use the binary neutron star parameters inferred for the short GRB 090227B, including the cosmological redshift $z = 1.61$. The comparison is made for both the dynamics given by the non-relativistic point-like particles approximation described in section \ref{sec:5.1} (dotted black curve) and the dynamics obtained from the EOB formalism, described in section \ref{sec:5.2}. In the case of the EOB approach, the radial potential $A(u;\nu)$ was calculated using post-Newtonian approximation (PN). The dotted-dashed black curve is $\rm A(u;\nu)=3PN$, using the Pad\`{e} approximant we calculated the $P^1_3[\rm A(u;\nu)= 3PN]$ (dashed black curve) and the $P^1_5[\rm A(u;\nu)=4PN]$ (solid black curve). The noise spectral density of Advanced LIGO, $S_h(f)$, is represented by the solid gray curve. Right panel: same as left panel but for a redshift $z = 0.09$, which would give a gravitational wave detection with $\langle{\rm SNR}\rangle=5$ by Advanced LIGO.}\label{fig:SNR}
\end{figure*}

In the left panel of Fig.~\ref{fig:SNR} we use the theoretically estimated redshift of GRB 090227B, $z=1.61$, which results in a contact frequency at the detector, $f_{\rm max}\approx 1534.19/(1+1.61)$~Hz$=587.81$~Hz. At such a redshift, $\langle {\rm SNR}\rangle \approx 0.32$ for Advanced LIGO, a much lower value than the threshold for a positive detection.

At this point, it is natural to ask at which distance the gravitational waves emission from the progenitor of GRB 090227B would have been detectable by the Advanced LIGO interferometer. We find that a $\langle {\rm SNR}\rangle=5$ would be produced if the GRB would be located at a redshift $z\approx 0.09$, namely at a distance $d_L\approx 381$~Mpc. We show in the right panel of Fig.~\ref{fig:SNR} the results for this hypothetical redshift for detection. In this case, the contact frequency at the detector is, $f_{\rm max}\approx 1534.19/(1+0.09)$~Hz$=1407.51$~Hz. The above numerical values are obtained for the most accurate case, $P^1_5[\rm A(u;\nu)=4PN]$.

We have discussed so far the issue of the conditions for a positive or negative detection by the Advanced LIGO interferometer of the spiraling phase of the progenitor of a genuine short GRB, as the binary neutron star progenitor of GRB 090227B. A different question are the conditions under which such a detection could give us reliable information on the structure of the neutron star components. In such a case, a much higher value of the SNR is needed. For instance, it has been recently \cite{2012PhRvD..85l3007D} estimated that a $\langle{\rm SNR}\rangle\approx 16$ is demanded to get knowledge of the neutron star equation of state, extracted from the tidal polarizability parameters of the binary components. This would imply a still shorter distance to the source that one derived above.

\section{Total energy output}\label{sec:7}

The gravitational waves emission dominates the energy loss during the spiraling phase while the X and Gamma rays dominate from the coalescence with the final emission of a short GRB if the total mass of the binary exceeds the critical mass for neutron star gravitational collapse. Thus, an upper limit for the gravitational wave emission radiated away can be obtained from the energy difference between the initial binary at time $t_0=0$ with separation $r_0$ and energy $E_0$, and the binary at time $t_f$ and separation $r_f=r_{AB,{\rm min}}=R_1+R_2$, with energy $E_{\rm f}$, when the two components touch each other. 

An absolute upper limit, for the gravitational wave energy emission, $\Delta E^{\rm max}_{\rm GW}$, can be therefore determined by the assumption of an infinite initial separation $r_0 \to \infty$, namely 
\begin{equation}
\Delta E^{\rm max}_{\rm GW}= \left|E_b(t_f) - E_b(t_0)\right|.
\end{equation}

For the neutron star binary discussed in this work for GRB 090227B, we obtain the absolute upper bound shown in Table \ref{tab:bounds}. The gravitational wave energy emission $\Delta E^{\rm max}_{\rm GW}$ which in the case of the genuinely short GRB 090227B is one order of magnitude smaller than the emitted electromagnetic energy $E^{\rm GRB}_{\rm tot}=2.83\times 10^{53}$ erg (see Table \ref{tab:GRB090227B}). 

\begin{deluxetable}{lccc}[!hbtp]
\tabletypesize{\scriptsize} 
\tablecaption{Upper limit for the total GW emission, $\Delta E^{\rm max}_{\rm GW}$ (erg).
\label{tab:bounds}}
\tablewidth{0pt}
\tablehead{
\colhead{Classic} &
\colhead{EOB $A$3PN } &
\colhead{EOB $P^1_3$[$A$3PN]} &
\colhead{EOB $P^1_5$[$A$4PN]} 
}
\startdata
$ 9.6\times 10^{52}$   &   $ 9.68\times 10^{52}$   & $ 7.41\times10^{52}$  & $ 7.42\times10^{52}$  
\enddata
\end{deluxetable}

It is also worth to mention that indeed this numerical value for $\Delta E^{\rm max}_{\rm GW}$ limits from above the results of full numerical integrations of the gravitational wave radiation emitted in the neutron star binaries during the entire process of spiraling and merging (see e.g.~\cite{2001A&A...380..544R}).

Additional contributions to the gravitational wave power due to higher multipole moments of the components such as angular momentum $J$ and quadrupole moment $Q$ (deformation) are conceptually relevant corrections to the above formulas (see e.g.~\cite{1995PhRvD..52.5707R} and references therein, for details); however they are quantitatively negligible for the present purpose. For instance, the first correction due to the spin angular momentum $J$ of the neutron star components is given by $-11/4\,j \Omega M$ in geometric units, where $j=c J/(G M^2)$ is the dimensionless angular momentum parameter. This correction is only of order $10^{-2}$ for a binary orbit of very high angular frequency $\sim$ kHz and for neutron stars with $M = 1.335 M_\odot$ and $j=0.4$. We recall that the fastest observed pulsar, PSR J1748-2246ad, has a rotation frequency of 716 Hz \citep{hessels06}, which gives $j\sim 0.51~I_{45}/(M_0/M_\odot)^2=0.26~I_{45}$ with the latter value for a canonical NS of $M=1.4 M_\odot$, $I_{45}$ is the moment of inertia in units of $10^{45}$ g cm$^2$. The first correction due to the quadrupole deformation multipole moment $Q$ of the neutron star, given by $-2~Q~\Omega^{4/3} M^{-5/3}$, is of order $10^{-3}$ for the same parameters with $Q \approx 4\times 10^{43}$~g~cm$^2\approx 3$~km$^3$, the latter value in geometric units.

\section{Conclusions}\label{sec:8}

We show that the observations of the genuinely short GRB 090227B lead to crucial information on the binary neutron star progenitor. The data obtained from the electromagnetic spectrum allows to probe crucial aspects of the correct theory of neutron stars and their equation of state. The baryon load parameter $B$ obtained from the analysis of GRB 090227B, leads to a most remarkable agreement of the baryonic matter expected to be ejected in a neutron star binary merger and validate the choice of the parameters of the binary components, $M_1 = M_2 = 1.34~M_\odot$, and $R_1=R_2=12.24$~km. This represents a test of the actual neutron star parameters described by the recent developed self-consistent theory of neutron stars \citep{2012NuPhA.883....1B} that takes into account the strong, weak, electromagnetic and gravitational interactions within general relativity and satisfy the condition of global charge neutrality.

We have discussed how the inference of the neutron star parameters, mass and radius, and the expected baryon load produced during the merger process, depends on the nuclear equation of state as well as on the condition of global and local charge neutrality. We have also argued that the current observational constraints of pulsars on the mass-radius relation of compact stars rule out an alternative scenario given by strange quark stars, although they have core-crust transition and crust properties similar to the ones of globally neutral neutron stars of \cite{2012NuPhA.883....1B}.

We computed the dynamics of the neutron star binary progenitor prior to the merger and emission of the GRB. We compare and contrast the classic description of the dynamics with the more general one given by the framework of the effective one-body formalism, which we use up to 4-PN order. We have shown that the classic binary dynamics overestimate the energy output in gravitational waves with respect to the more accurate dynamics of the effective one-body formalism. In addition, we showed the necessity of using the Pad\`e approximant in order to keep the solution stable close to the merger point.  

We estimate the detectability of GRB 090227B by the Advanced LIGO interferometer, by computing the SNR up to the contact point of the binary components, for the theoretically inferred cosmological redshift, $z=1.61$ \citep{2013ApJ...763..125M}; see left panel of Fig.~\ref{fig:SNR}. We find that at such a redshift, the gravitational waves signal would produce a $\langle{\rm SNR}\rangle\approx 0.32$, a value lower than the one needed for a positive detection, $\langle{\rm SNR}\rangle= 5$. We turn to estimate the redshift at which Advanced LIGO would detect this GRB with a $\langle{\rm SNR}\rangle=5$ (see Fig.~\ref{fig:SNR}, right panel) we obtained $z\approx 0.09$ or a distance to the source $d_L\approx 381$~Mpc. Unfortunately, in the last 40 years, no such a GRB has been observed.

From the dynamics, we estimated the total energy release in form of gravitational waves up to the point where the stars touch each other (see Table \ref{tab:bounds}); we compare and contrast it with the energy in X and Gamma rays released in the final emission of the GRB. We conclude that the emission of X and Gamma rays in a short GRB by a binary neutron star merger is at least one order of magnitude larger than the gravitational wave emission in the entire life of the binary including the last plunge.

\acknowledgements{We thank the anonymous referee for the comments and suggestions which helped to improve the presentation of our results.} F.~G.~Oliveira acknowledges the support given by the International Relativistic Astrophysics Erasmus Mundus Joint Doctorate Program under the Grant 2012–-1710 from EACEA of the European Commission.


%
\end{document}